# Application of a New Designed ROADM to Improve the Performance of Elastic Optical Network

Faranak Khosravi
*Dep. Electrical Engineering*
*University of Texas at San Antonio*
San Antonio, USA
faranak.khosravi@my.utsa.edu

Subbulakshmi Easwaran
*Dep. Electrical Engineering*
*University of Texas at San Antonio*
San Antonio, USA
subbulakshmi.easwaran@my.utsa.edu

Mehdi Shadaram
*Dep. Electrical Engineering*
*University of Texas at San Antonio*
San Antonio, USA
mehdi.Shadaram@utsa.edu

*ABSTRACT: The quick rise of emerging technologies like spectrally efficient multicarrier with a higher order modulation and bandwidth variable wavelength selective switches have caused a shift of the optical network architecture from fixed to the flexible elastic optical network (EON), in terms of bit rate, center frequency spacing, modulation format, and optical reach. By using bandwidth-variable transceivers, colorless flexible-grid reconfigurable optical add-drop multiplexers (ROADM), as well as a choice of different optical source types, operators can provide a range of service types and improve network efficiency. Implementation of the EON network requires careful consideration, especially in terms of the ROADM structure to achieve optimal performance. We propose a low-cost ROADM structure with an order-based connection method that includes a high degree with a blocking probability of less than $10^{-4}$. Then, we applied the new design of ROADM in EON that transported 1 Tb/s. In the final step, we conclude that our system accommodated 20 percent more traffic demand over an average of the low-rate wavelength approach.*

## I. INTRODUCTION

The International Telecommunication Union (ITU-GT)'s 694.1 guideline was modified in 2012 with the intention of introducing the flexible grid option in accordance with a novel notion of frequency slot to affirm elastic optical networks (EON)s spectrum allocation approach [1]. As an integral multiple of 12.5 GHz is used to define the slot width in EON, the spectral efficiency (SE) will improve by 56% if the 32 GB, symbol rate and the zero roll-off Nyquist filtering are used [2]. Finer granularity meets the hardware requirements of reconfigurable optical add-drop multiplexers (ROADM) and the management system with a small drop in the SE for static traffic demands. The wavelength-selective switch (WSS), which shifts the wavelengths from an input port to an N output port in ROADMs, is regarded as the fundamental requirement for achieving total bandwidth flexibility [3, 4]. Increased traffic and use of additional fibers on each link of an optical network necessitates the use of higher degree ROADMs, which allow for flexibility in the add/drop rate in addition to allowing transmission in a variety of directions [5]. Some studies have looked into high-degree ROADMs based on commercial small-size WSSs because it is challenging to fabricate large-size WSSs [5, 6]. Instead of adding a new chassis to improve ROADM capacity, the implementation of the existing chassis for future scaling is considered the key to minimize the cost [5, 6]. In this work, a new model of a low cost and higher degree three-stage switching Flex ROADM was proposed for the first time to the best of our knowledge to assess the impact of EON network requirements and then compare it with the performance of ROADM that was designed in [7] for wavelength division multiplexing (WDM) system. The work is organized as follows: a methodology that includes discussion of ROADM structures, and its applications in EON are presented in section II and finally, the results and conclusion are provided in section III.

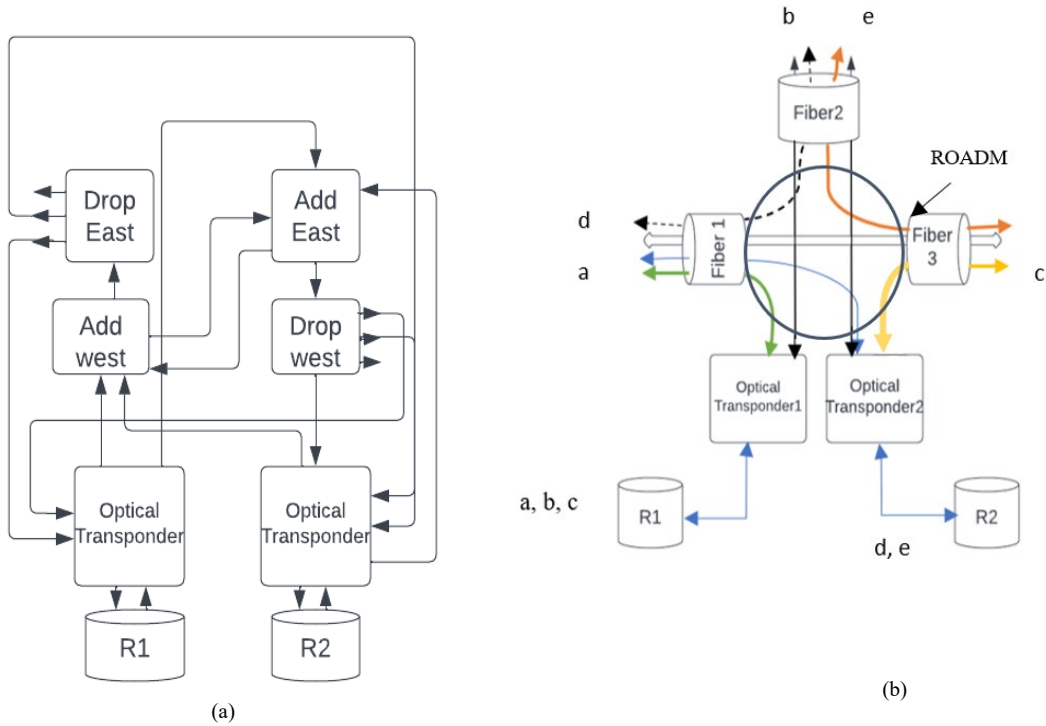

Fig1. Structure of ROADM (a) node architecture; b) ROADM node-to-multipoint connections.

## II. METHODOLOGY

Based on designing ROADM in [8], Fig1 shows the architecture and operation of an optical transponder-equipped ROADM. According to Fig1, traffic flows a, b, c, d, and e, respectively, were present in the data traffic coming from the interfaces of client routers 1 and 2. At optical transponders 1 and 2, each traffic flow was converted into an optical flow with a unique wavelength. Several optical connections were subsequently formed from a single node ROADM, as seen in Fig. 1 (a). Fig. 1 (b) shows the output of ROADM applied to three separate optical fibers. Each output side ROADM transmits the desired optical flows according to the predetermined path. The optical flows, a, b, and c are output from client router 1 through fibers 1, 2, and 3 respectively. As a result, Fig. 1 demonstrates a single optical transponder created numerous optical connections. To enhance the performance of ROADM compared to the approach proposed in [8], our research focused on increasing the degree of ROADM to provide high network switching capacity and address the issue of WSS's large size by reducing it. The cluster ROADM node comprises three chassis, namely the line, add/drop, and interconnect chassis. Our research proposed a low-cost ROADM cluster node that utilizes pre-existing chassis for improved reusability and flexibility. To accommodate the N incoming and outgoing fibers of the cluster node, each line chassis has M connection cards and N line cards, and it can support a maximum of M+N WSS cards. The slots of the add/drop chassis are divided between add/drop cards and interconnect cards. The ROADM cluster node consists of M interconnect chassis that links E (number of line chassis) × line chassis and F (number of add/drop chassis) × add/drop chassis. Additionally, a cluster controller is set up to manage the operations of the cluster ROADM node by communicating with each chassis controller [7]. Moreover, for connecting line chassis and add/drop chassis, each interconnect chassis has S interconnect cards, where S=E+F. The ROADM cluster node has a total degree of E×N. Moreover, the new design of ROADM uses nonblocking architecture to provide high-capacity network. This means that our design ROADM can reconfigure its optical paths to allow new connections or changes to existing connections without affecting the transmission of other signals that are already passing through the system. It is rearrangeable nonblocking if k (the number of middle switches) is greater than or equal to n (the number of outlets in each third stage array or the number of inlets in each first stage array), which implies that in the absence of rearrangements, the network is blocking for n<k< 2n-1. It was suggested that using N <M <1.2×N because M < 2n-1 choice results in more common equipment costs and fewer degrees. A cluster node, which is responsible for tasks of ROADM such as monitoring the network performance, managing the optical switch configurations, can be scaled to 224 degrees in comparison to the design 224 provided in [7] by choosing N=14, M=16, and assuming F= 0.

We used blocking probability in our system to measure the system's ability to handle incoming traffic without causing delays or interruptions. Blocking probabilities can be computed by using Lee's approach for random routing under the assumption that incoming traffic is spread equally over M links. Equation (1) shows the total blocking probability of ROADM in our system. The parameter 'd' in equation (1) is a packing degree (and it is a mathematical notion, does not exert any deterministic control over the switch) and in this research was assuming 320 wavelength ITU 12.5 GHz grid for EON system and is the packing degree for wavelength ω. In our simulation, we considered N=14, M=16, S=16, parameter 'a' as such that carried traffic per line cards, and $Λ_d$ that offered light path traffic load on degree d. The total blocking probability is given by the following expression:

$$P_b = [1 - (1 - \frac{Na-d}{M-d})^2]^{(M-d)}, d=0,...,Na \quad (1)$$

$$P_{bd} = \frac{\sum_d P_b \cdot Λ_d}{\sum Λ_d} \quad (2)$$

As order-based connection methods were chosen, the average blocking can be obtained as follows:

$$P_b^d = \frac{1}{48000} \sum_{i=0}^{48000} P_{bd}(p_i),$$
$$p_i = 0, \frac{1}{48000}, \frac{2}{48000}, ...., 1 \quad (3)$$

$P_{bd}$ was calculated from equations 2, 3 and d is a packing degree. The limitation of sigma from 0 to 48000 was chosen because our ROADM was applied in EON in full load simulation of 48000 connections in the simulation. In previous designs, the IP over low-rate wavelength approach required a large number of router interfaces on the client side and numerous optical transponders and add/drop ports on the network side. To address this issue, we conducted simulations of a newly designed ROADM node, which

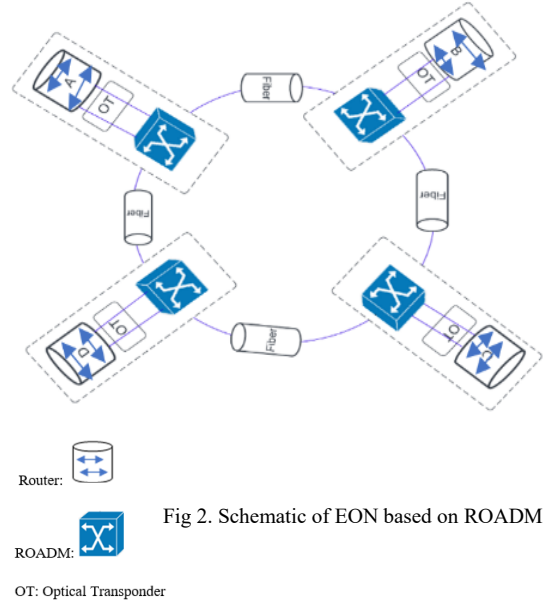

Fig 2. Schematic of EON based on ROADM.

Router:
ROADM:
OT: Optical Transponder

was subsequently applied in EON test networks to evaluate its performance. Our designed network, depicted in Fig. 2, is more scalable and effective compared to other networks [4-7]. We assumed the use of a 1Tb/s router interface and a 1Tb/s multi-flow optical transponder to link a client router to an elastic optical path network. Our multi-flow transponder converts client traffic into two optical flows of 400 Gbps each and one optical flow of 200 Gbps. Our newly designed ROADMs enabled a client flow to be converted into a single super channel optical flow, which can be carried over the route, making the EON system more efficient. This structure can provide significant spectrum savings when combined with the elastic optical path network's variable channel spacing.

### III. RESULTS

In this section, we analyze the performance of the proposed ROADM cluster node with five scenarios as listed in Table 1. We simulated the system under full load with random wavelength connectivity with an order-based connection method. Case 1 had a cluster ROADM node with 112 degrees and 100% add/drop capacity, Cases 2 to 4 had add/drop rates that range from 14% to 60%, and finally, Case 5 considered as a cluster ROADM node with 224 degrees and no add/drop connectivity. Each scenario was performed through a full-load simulation with 48000 wavelengths connections per connectivity map used to calculate the blocking rate. To evaluate the performance of our proposed high-degree ROADM architecture, we first considered the impact of uniform traffic load on each degree. We observed that the blocking probabilities of all the ROADMs increased with the traffic load on each degree, as more light path requests needed to be set up, leading to a higher chance of being blocked. Comparing the two interconnection patterns, we found that our proposed interconnection pattern achieved better blocking performance than the random interconnection pattern. This was because our proposed interconnection pattern ensured that each add/drop module was connected to at least one fiber link degree on any ROADM degree, while the random interconnection pattern did not adopt this strategy, resulting in degraded performance.

Table 1. Summary of fully loaded cluster node simulation modeling for N=14 and M=16 for 5 different cases

|  | Case 1 | Case 2 | Case 3 | Case 4 | Case 5 |
|---|---|---|---|---|---|
| Number of Line Chassis (E) | 8 | 10 | 12 | 14 | 16 |
| Number of add/drop (F) | 8 | 6 | 4 | 2 | 0 |
| Number of degrees N× E (here we consider N is 14) | 112 | 140 | 168 | 196 | 224 |
| Add/ Drop rate | 100% | 60% | 33% | 14% | 0 |
| Probability blocking | $3\times10^{-6}$ | $4.5\times10^{-6}$ | $6.8\times10^{-6}$ | $0.9\times10^{-5}$ | $1.1\times10^{-5}$ |

Table 2. Evaluation results of spectral resource utilization efficiency assumed for each optical flow.

| Bit rate | | 40 (Gb/s) | 100(Gb/s) | 400(Gb/s) | 1 (Tb/s) |
|---|---|---|---|---|---|
| Spectral Width (GHz) | EON Original [4] | 25 | 50 | 100 | 150 |
| | New design EON | 25 | 45 | 90 | 130 |

We also compared our designed ROADM in EON with the ROADM structure used in WDM. Interestingly, we found that as the number of wavelengths in each fiber increased from 80 to 320, the difference in performance between the two architectures decreased. In situations where wavelength resources are insufficient, we found that EON's flexibility in using an order-based connecting

We confirmed that the results obtained from our analytical model were in close agreement with those from the simulation. Table 2 presents the evaluation results of the spectral resource utilization efficiency for each optical flow. We assumed a multi-flow optical transponder with 10 subchannel transmitters, each capable of producing 10 separate optical flows at a rate of 100 Gb/s. For the IP over low-rate wavelength technique we assumed 100-Gb/s wavelengths, and made individual requests by randomly selecting a pair of routers and choosing a capacity between 40 and 1Tb/s in steps of 20 Gb/s (although only four important bit rates are mentioned in the table). Compared to other IP over low-rate wavelength strategies [4], our proposed IP over elastic wavelength approach can support an additional 20% of traffic demand on average, resulting in significant increase in efficiency and cost reductions for optical networks with multi-flow requirements. Therefore, we conclude that our proposed ROADM has better performance and is well-suited for use in EON.

# CONCLUSION

In this article, we propose a novel ROADM architecture and apply it to an EON composed of a multi-flow optical transponder. Our proposed three-stage ROADM cluster node is highly scalable, adaptable, and efficient. It simplifies the WSSs significantly and offers better than $10^{-4}$ blocking at a full load of 48,000 connections. We construct the suggested cluster node by using an existing ROADM chassis and minimizing the cost. Then, we demonstrate how per-channel customization can increase the network capacity when using ROADM with elastic transponder technology.

# REFRENCESS